\documentclass[apj]{emulateapj}

\usepackage{natbib}
\usepackage{color}

\usepackage{bm}
\usepackage{lscape}

\newcommand{\eg}{e.g., }
\newcommand{\ie}{i.e., }

\newcommand{\kms}{km~s$^{-1}$}

\newcommand{\Nifs}{$^{56}$Ni}

\def\gsim{\mathrel{\rlap{\lower 4pt \hbox{\hskip 1pt $\sim$}}\raise 1pt
\hbox {$>$}}}
\def\lsim{\mathrel{\rlap{\lower 4pt \hbox{\hskip 1pt $\sim$}}\raise 1pt
\hbox {$<$}}}

\def\ion#1#2{{\rm #1}~{\sc #2}}

\shorttitle{3D Explosion Geometry of Stripped-Envelope Core-Collapse SNe I}
\shortauthors{Tanaka et al.}

\begin{document}

\title{
Three-Dimensional Explosion Geometry of Stripped-Envelope Core-Collapse Supernovae. I. Spectropolarimetric Observations
\altaffilmark{1}}
\author{
Masaomi Tanaka\altaffilmark{2,3}, 
Koji S. Kawabata\altaffilmark{4},
Takashi Hattori\altaffilmark{5},
Paolo A. Mazzali\altaffilmark{6,7}, 
Kentaro Aoki\altaffilmark{5},
Masanori Iye\altaffilmark{2},
Keiichi Maeda\altaffilmark{3},
Ken'ichi Nomoto\altaffilmark{3},
Elena Pian\altaffilmark{8},
Toshiyuki Sasaki\altaffilmark{5}, and
Masayuki Yamanaka\altaffilmark{4,9}
}

\altaffiltext{1}{Based on data collected at Subaru Telescope, 
which is operated by the National Astronomical Observatory of Japan.}
\altaffiltext{2}{National Astronomical Observatory, Mitaka, Tokyo, Japan; masaomi.tanaka@nao.ac.jp}
\altaffiltext{3}{Institute for the Physics and Mathematics of the Universe, University of Tokyo, Kashiwa, Japan}
\altaffiltext{4}{Hiroshima Astrophysical Science Center, Hiroshima University, Higashi-Hiroshima, Hiroshima, Japan}
\altaffiltext{5}{Subaru Telescope, National Astronomical Observatory of Japan, Hilo, HI}
\altaffiltext{6}{Max-Planck Institut f\"ur Astrophysik, Karl-Schwarzschild-Strasse 2 D-85748 Garching bei M\"unchen, Germany}
\altaffiltext{7}{Istituto Naz. di Astrofisica-Oss. Astron., vicolo dell'Osservatorio, 5, 35122 Padova, Italy}
\altaffiltext{8}{Istituto Naz. di Astrofisica-Oss. Astron., Via Tiepolo, 11, 34131 Trieste, Italy}
\altaffiltext{9}{Department of Physical Science, Hiroshima University, Higashi-Hiroshima, Hiroshima, Japan}

\begin{abstract}
We study the multi-dimensional geometry of supernova (SN) explosions  
by means of spectropolarimetric observations of stripped-envelope SNe, 
\ie SNe without a H-rich layer.
We perform spectropolarimetric observations of 2 stripped-envelope SNe, 
the Type Ib SN 2009jf and the Type Ic SN 2009mi. 
Both objects show non-zero polarization at the wavelength of the strong lines.
They also show a loop in the Stokes $Q-U$ diagram, 
which indicates a non-axisymmetric, 
three-dimensional ion distribution in the ejecta.
We show that five out of six stripped-envelope SNe
which have been observed spectropolarimetrically so far
show such a loop. This implies that 
a three-dimensional geometry is common in stripped-envelope SNe.
We find that stronger lines tend to show higher polarization. 
This effect is not related to the geometry,
and must be corrected to compare the polarization
of different lines or different objects.
Even after the correction, however, there remains a dispersion of 
polarization degree among different objects.
Such a dispersion might be caused by three-dimensional clumpy ion distributions
viewed from different directions.
\end{abstract}


\section{Introduction}
\label{sec:intro}

\subsection{Explosion Geometry of Supernovae}
\label{sec:intro_SN}

Core-collapse supernovae (SNe) are the explosions of massive 
stars at the end of their lives. 
Besides their importance as the end points of stellar evolution,
SNe play an important role on chemical and dynamical evolution
of galaxies.
However, the explosion mechanism of SNe has been unclear 
for a long time after the first concept by \citet{burbidge57}
and the first numerical simulation by \citet{colgate66}.

Modern detailed numerical simulations agree 
in that a successful explosion cannot be obtained in 
one-dimensional simulations
\citep{rampp00,liebendorfer01,thompson03,sumiyoshi05}.
Recently, more and more attention is being paid to multi-dimensional effects, 
such as convection \citep[\eg][]{herant94,burrows95,janka96}
and Standing Accretion Shock Instability 
\citep[SASI, \eg][]{blondin03,scheck04,ohnishi06,foglizzo07,iwakami08}.
In fact, some successful explosions have  been obtained in 
two-dimensional (2D) simulations \citep{buras06,marek09,suwa10}.

Thanks to the advance of numerical simulations, 
some simulations have been done in fully three-dimensional (3D) space
\citep{nordhaus10,wongwathanarat10,takiwaki12,kuroda12}.
Interestingly, \citet{nordhaus10} show that it is easier to 
obtain a successful explosion in 3D simulation than in 2D simulations
\citep[but see also][]{hanke11}.

Given these circumstances, it is important to obtain
observational constraints on the multi-dimensional geometry of SN explosions.
The most direct way is the observation of 
spatially resolved supernovae.
\citet{isensee10} and \citet{delaney10} performed detailed observations 
of Galactic supernova remnant Cassiopeia A.
Using a combination of imaging and spatially resolved spectroscopy, 
they reconstructed the 3D geometry of the SN. 
Similarly, direct imaging \citep{wang02} 
and spatially resolved spectroscopy \citep{kjaer10}
have been performed for SN 1987A in the Large Magellanic Cloud.
However, the number of objects suitable for such observations is limited.

Observations of extragalactic SNe can also provide hints of 
explosion geometry, although at this distance SNe are point sources.
For example, spectroscopic observations of SNe at
$>$1 yr after the explosion provide information of explosion geometry.
At such late epochs, SN ejecta are optically thin.
Therefore, this method is sensitive to the geometry of the innermost ejecta.
By analyzing line profiles of nebular emission, 
bipolar geometry of SNe has been suggested
(\eg \citealt{mazzali01,maeda02,mazzali05,maeda08,modjaz08,tanaka0908Dneb,taubenberger09,maurer10},
but see also \citealt{milisavljevic10}).

Polarization at early phases ($\lsim$ 50 d after the explosion)
is another powerful method to 
study the multi-dimensional geometry of extragalactic SNe 
\citep[see][for a review]{wang08}.
In contrast to late phase spectroscopy, polarization is sensitive to
the geometry of the outer ejecta.
In particular, spectropolarimetry can probe 
3D geometry, as explained in the following section.
In this paper, we study the multi-dimensional explosion 
geometry of SNe using spectropolarimetric observations.

\subsection{Power of Spectropolarimetry}
\label{sec:intro_specpol}

\begin{figure}
\begin{center}
\includegraphics[scale=0.4]{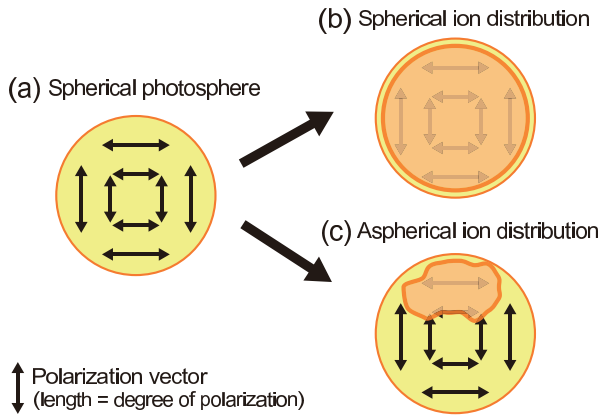}
\caption{Schematic illustration of polarization in the SN ejecta
(see also \citealt{kasen03,leonardfilippenko05,wang08}).
(a) When the photosphere is spherical, 
polarization is cancelled out, and no polarization is expected. 
At the wavelength of a line, polarization produced by 
the electron scattering is depolarized by the line transition.
(b) When the ion distribution
is spherical, the remaining polarization is cancelled, 
and no polarization is expected.
(c) When the ion distribution is not spherical, 
the cancellation becomes incomplete, and 
line polarization could be detected
\citep[\eg][]{jeffery89,kasen03,wang08}.  
\label{fig:ejecta}}
\end{center}
\end{figure}

\begin{figure*}
\begin{center}
\begin{tabular}{ccc}
\includegraphics[scale=0.30]{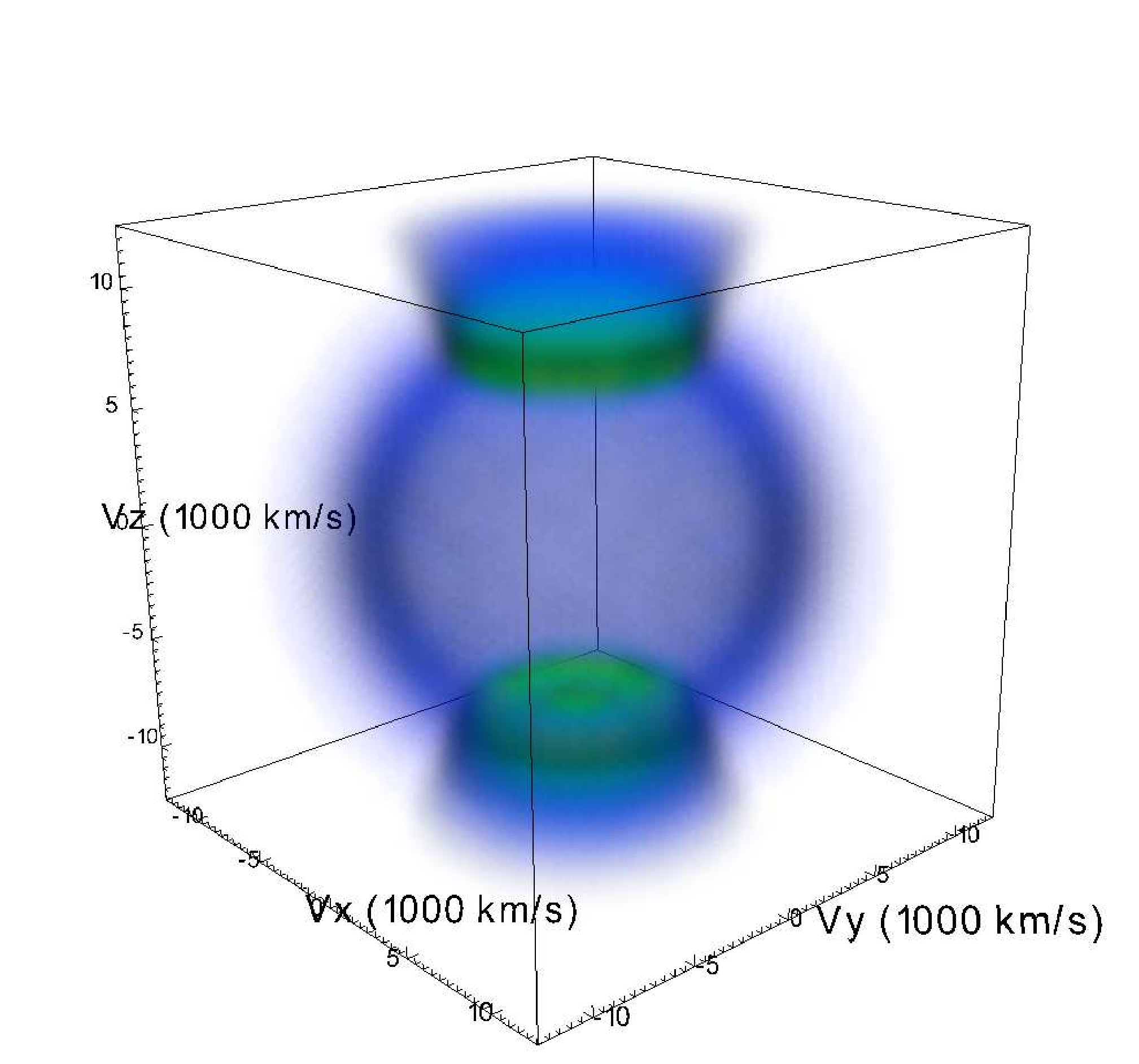} &
\includegraphics[scale=0.5]{f2b.eps} &
\includegraphics[scale=0.50]{f2c.eps} \\
\includegraphics[scale=0.30]{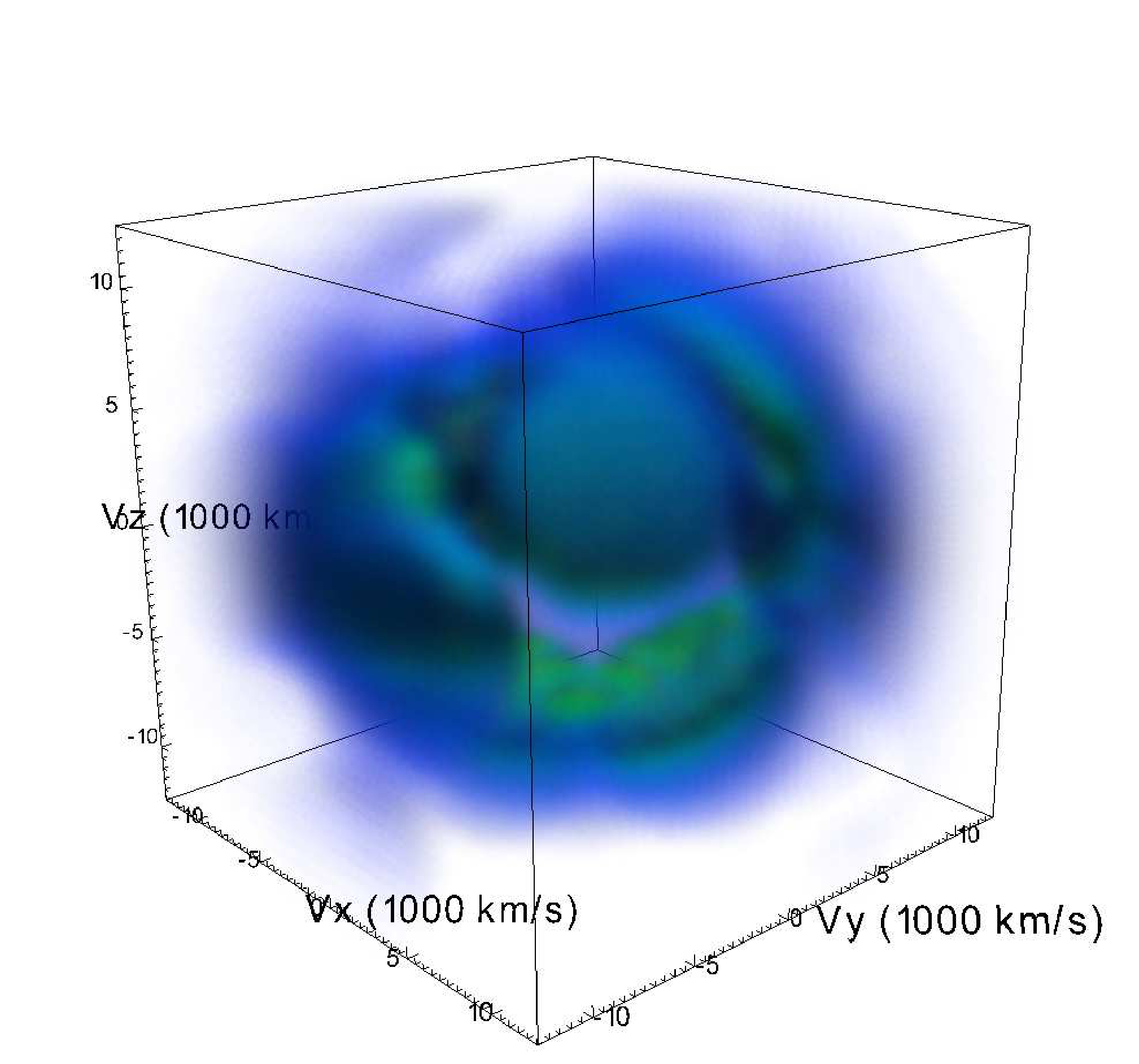} &
\includegraphics[scale=0.5]{f2e.eps} &
\includegraphics[scale=0.50]{f2f.eps}
\end{tabular}
\caption{
Simulated line polarization for a 2D bipolar model ({\it top}) 
and 3D clumpy model ({\it bottom}).
({\it Left}) The distribution of the line optical depth.
({\it Middle}) The total flux spectrum and polarization spectrum.
The polarization angle is consistent with a constant in the 2D model, 
while it changes across the line in the 3D model.
({\it Right}) The same simulated polarization but in the $Q-U$ diagram.
Different colors represent different Doppler velocity, 
according to the color bar above the plots.
The polarization data show a straight line (constant angle) in the 2D model,
while the data show a loop (variable angle) in the 3D model.
Details of the simulations will be given in a forthcoming paper.
\label{fig:intro_sim}}
\end{center}
\end{figure*}

Polarization in SNe is generated by electron scattering.
Since the polarization degree of the scattered light 
becomes maximum when the scattering angle is $90^{\circ}$, 
the expected polarization map for a spherical photosphere 
looks like (a) in Figure \ref{fig:ejecta}.
However, SNe are observed as point sources, and thus
all the polarization vectors cancel out, 
resulting in no polarization.
\footnote{
When the photosphere is deformed,
cancellation of polarization becomes incomplete.
As a result, continuum polarization will be detected 
\citep{shapiro82,hoeflich91}.
Although continuum polarization is difficult to 
distinguish from interstellar polarization,
multi-epoch observations have revealed that 
there is non-zero continuum polarization in Type IIP SNe 
after the plateau phase \citep{leonard06,chornock10IIP}.
}

Absorption and re-emission by a bound-bound transition,
which make a P-Cygni profile in the expanding ejecta, 
tend to cause depolarization \citep{jeffery89,kasen03}.
Thus, the polarization produced by the electron scattering is 
effectively concealed by the interaction with lines.
If an ion is distributed spherically (b), 
polarization vectors are still cancelled out,
and no polarization is expected at the wavelength of the line.
Polarization would be detected 
at the wavelength of the absorption line 
only if the ion is distributed asymmetrically (c),
because of the incomplete cancellation.
Thus, we can identify an asymmetric ion distribution by 
detecting line polarization.

Note that an asymmetric ion distribution is expected either from 
an asymmetric element distribution,
which could be produced by a multi-dimensional explosion 
\citep[\eg][]{nagataki97,kifonidis00,kifonidis03,maeda03,tominaga09,fujimoto11},
or by an asymmetric ionization/temperature structure, 
which is caused by an asymmetric distribution 
of \Nifs\ \citep{tanaka07}.

The properties of linear polarization are fully described 
by the Stokes parameters.
Throughout this paper, we define the Stokes parameters as
a fraction of the total flux: $Q \equiv \hat{Q}/I$ and 
$U \equiv \hat{U}/I$, where $\hat{Q}$ and $\hat{U}$ can be 
expressed by $\hat{Q} = I_{0} - I_{90}$ and
$\hat{U} = I_{45} - I_{135}$, respectively. 
Here $I_{\theta}$ is the intensity 
measured through the ideal polarization filter with an 
angle $\theta$.
From the Stokes parameters $Q$ and $U$,
the polarization position angle $\theta$ is obtained by 
\begin{equation}
2\theta = {\rm atan}(U/Q).
\end{equation}
The position angle is measured from north to east on the sky.

Figure \ref{fig:intro_sim} shows the expected line polarization
in a 2D bipolar model (top) and a 3D clumpy model (bottom).
The polarization spectra are calculated by Monte Carlo method
for arbitrary line optical depth distributions,
under assumptions similar to those in \citet{kasen03,hole10}.
The details of the simulations will be given in a forthcoming paper.
The left panels show the distribution of line optical depth.
In the 2D model, the opacity is enhanced by a factor of 10 
in the polar region, while in the 3D model, 
the opacity is similarly enhanced in the randomly distributed clumps.

In both cases, the polarization changes at the wavelength of the 
line (middle panels), 
which is caused by the asymmetric distribution of line optical depth.
However, the behavior of the polarization angle is different.
In the 2D case, the polarization angle is constant 
(some fluctuation is caused by the Monte Carlo noise),
while in the 3D case, it largely changes across the line.
Since the SN ejecta expand homologously and 
the velocity can be used as a radial coordinate,
a change in the polarization angle means 
a radially-dependent ion distribution.

This effect is more clearly seen in the $Q-U$ plane.
The right panels in Figure \ref{fig:intro_sim} show
the same simulated polarization spectrum in the $Q-U$ plane.
Different colors represent different Doppler velocity, 
according to the color bar above the plots.
The polarization for the 2D model shows a straight line 
in the $Q-U$ plane, while that for the 3D model
shows a ``loop'', which corresponds to a change in the 
polarization angle.

Therefore, a change in the polarization angle or 
the loop in the $Q-U$ plane is indicative of a 3D ion distribution.
In order to fully utilize the power of spectropolarimetry, 
we should obtain high-quality data
to discriminate such a feature from noise.
Loops in the $Q-U$ plane have been commonly observed 
in Type Ia SNe \citep{wang0301el,kasen03,chornock08,patat09}
and in some core-collapse SNe 
\citep[\eg][]{maund0705bf,maund0701ig,maund09}.

\subsection{This Paper}

In this paper, we show our 
spectropolarimetric observations of 2 Type Ib/c SNe with Subaru telescope
(Sections \ref{sec:obs} and \ref{sec:results}).
The first target is the 
Type Ib SN 2009jf (discovered in NGC 7479 by \citealt{li09} 
on UT 2009 September 27.33, and classified by \citealt{kasliwal09,sahu09}). 
See \citet{sahu11} and \citet{valenti11} for more details.
The other target is the Type Ic SN 2009mi 
(discovered in IC 2151 by \citealt{monard09} 
on UT 2009 December 12.91, and classified by \citealt{kinugasa09}).

We then summarize the spectropolarimetric data of stripped-envelope SNe 
published so far (Section \ref{sec:properties}).
We argue that a non-axisymmetric, 3D geometry is common 
in stripped-envelope SNe.
We find a relation between line polarization and the depth of absorptions,
\ie a stronger line tends to show a higher line polarization.
Even after correcting for the effect of the absorption depth, 
there remains a dispersion in the polarization degree among different objects.
The implications of this dispersion are discussed.
Finally, we give conclusions in Section \ref{sec:conclusions}.

\begin{deluxetable*}{lccccc} 
\tablewidth{0pt}
\tablecaption{Log of Observations}
\tablehead{
Object &
Date (UT) & 
Date (MJD) & 
Exposure time (s)&
Airmass &
Comment 
}
\startdata
 SN 2009jf ($+9.3$ d)   & 2009 Oct 24.3  & 55128.3 & $(600 \times 4) \times 6 $         & 1.03 -- 1.43 & SN                    \\
 BD+28$^{\circ}$4211     & 2009 Oct 24.2  & 55128.2 & $(20 \times 4) + (40 \times 4)$    & 1.01         & unpolarized/flux std. \\
 G191-B2B               & 2009 Oct 24.6  & 55128.6 & $(60 \times 4) \times 2$           & 1.33         & unpolarized std.      \\ 
 Hiltner 960            & 2009 Oct 24.2  & 55128.2 & $20 \times 4$                      & 1.08         & polarized std.        \\  \hline
 SN 2009mi ($+26.5$ d)  & 2010 Jan 8.3   & 55204.3 & $(600 \times 4) + (1000 \times 4)$ & 1.26 -- 1.83 & SN                    \\
 G191-B2B               & 2010 Jan 8.2   & 55204.2 & $(60 \times 4) \times 2$           & 1.43         & unpolarized/flux std. \\
 HD 14069               & 2010 Jan 8.2   & 55204.2 & $(5 \times 4) \times 2$            & 1.03         & unpolarized std.      \\
 HD 251204              & 2010 Jan 8.2   & 55204.2 & $20 \times 4$                      & 1.48         & polarized std.        
\enddata
\tablecomments{
All the observations were performed with a $0.8''$ width offset slit,
a 300 lines mm$^{-1}$ grism, and the Y47 filter, giving the wavelength coverage 4700-9000 \AA\
and the wavelength resolution $\Delta \lambda \simeq $ 10 \AA.}
\label{tab:obslog}
\end{deluxetable*}

\section{Observations}
\label{sec:obs}

We have performed spectropolarimetric observations of SNe 2009jf and 2009mi 
with the 8.2m Subaru telescope equipped
with the Faint Object Camera and Spectrograph \citep[FOCAS,][]{kashikawa02}
on UT 2009 October 24.3 (MJD=55128.3) 
and 2010 January 8.3 (MJD=55204.3), respectively.
These epochs correspond to $t=+9.3$ and $+26.5$ days from the $B$ band
maximum (MJD=55118.96 for SN 2009jf according to \citealt{sahu11}, 
and MJD = 55177.8 for SN 2009mi, based on our observations). 
Hereafter, $t$ denotes the days after the $B$-band maximum.
The log of observations are shown in Table \ref{tab:obslog}.

For both observations,
we used an offset slit of $0.8''$ width,
a 300 lines mm$^{-1}$ grism, and the Y47 filter.
This configuration gives a wavelength coverage of 4700-9000 \AA.
The wavelength resolution is $\Delta \lambda \simeq $ 10 \AA.
For the measurement of linear polarization, 
we use a rotating superachromatic half-wave plate 
and a crystal quartz Wollaston prism.
One set of observations consists of
the integrations with 
$0^{\circ}, \ 45^{\circ},\ 22.5^{\circ}$, and $67.5^{\circ}$ 
positions of the half-wave plate.
From this one set of exposures,
Stokes parameters $Q$ and $U$ are derived
as described by \citet{tinbergen96}.

For the observations of SN 2009jf 
we performed six sets of the integrations at the four angles
with a total exposure time 4.0 hr.
For SN 2009mi, we performed five sets of the integrations at the four angles
with a total exposure time of 3.8 hr.
Typical seeing during the observations was
$0.8"$ and $1.1-1.5"$ for SNe 2009jf and 2009mi, respectively.

After deriving the Stokes parameters for 
each set of observations, these were combined.
Then, the instrumental polarization
($\sim$ 0-0.4 \%) was evaluated and subtracted using 
unpolarized stars \citep{schmidt92pol};
G191-B2B (for both objects),
BD+28$^{\circ}$4211 (for SN 2009jf)
and HD 14069 (for SN 2009mi) observed on the same night.
The reference axis of the position angle was 
calibrated by the observation of the strongly polarized star 
Hiltner 960 \citep[for SN 2009jf]{schmidt92pol} and
HD 251204 \citep[for SN 2009mi]{turnshek90}.
The wavelength dependence of the optical axis of the half-wave plate
was corrected using the dome flat taken through a fully-polarizing filter.
The total flux was calibrated using the observation 
of the spectrophotometric standard stars BD+28$^{\circ}$4211 
and G191-B2B \citep{massey88,massey90,oke90} 
for SNe 2009jf and 2009mi, respectively.
Telluric absorption lines were also removed using 
the spectrum of these standard stars.

\begin{figure*}
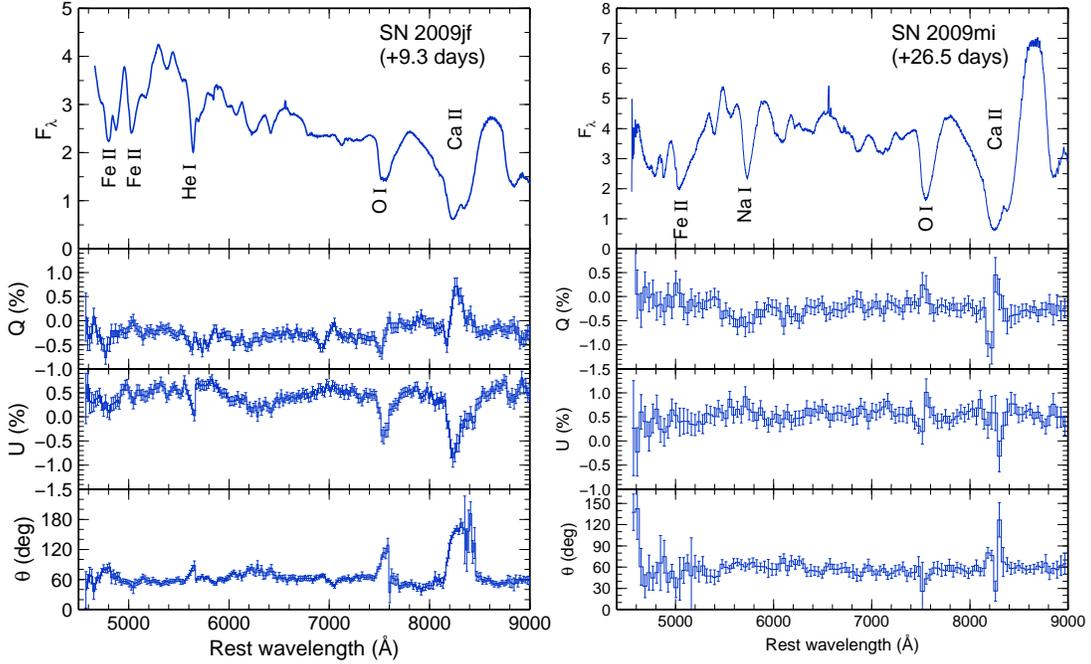

\begin{center}
\begin{tabular}{cc}
\includegraphics[scale=0.40]{f3a.eps} &
\includegraphics[scale=0.40]{f3b.eps} 
\end{tabular}
\caption{
Total flux spectrum and polarization spectrum of 
SNe 2009jf (left) and SN 2009mi (right).
The total flux is shown in units of 
$10^{-15}$ and $10^{-16}$ ${\rm \ erg\ s^{-1}\ cm^{-2}\ \AA}$ 
for SNe 2009jf and 2009mi, respectively.
The polarization data are binned into 25 and 50 \AA\ 
for SNe 2009jf and 2009mi, respectively.
In the plot of Stokes parameters and polarization angle,
contribution of the interstellar polarization is {\it not} corrected for.
\label{fig:pol}}
\end{center}
\end{figure*}

\begin{figure*}
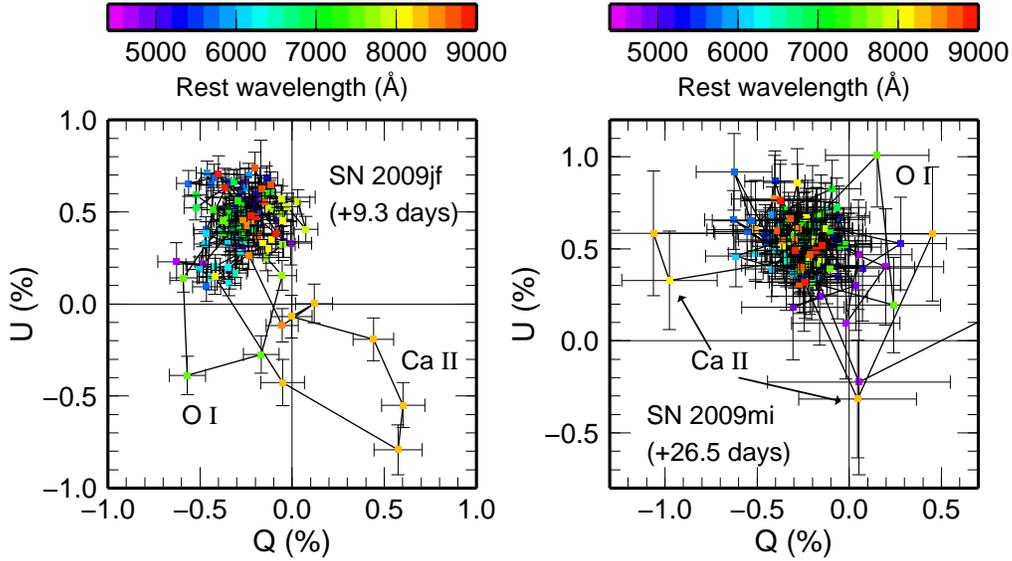

\begin{center}
\begin{tabular}{cc}
\includegraphics[scale=0.70]{f4a.eps} &
\includegraphics[scale=0.70]{f4b.eps}
\end{tabular}
\caption{
Observed polarization data of SNe 2009jf (left) and 2009mi (right)
in the $Q-U$ plane.
The polarization data are binned into 50 \AA.
The interstellar polarization is {\it not} corrected for.
Different colors show different wavelengths as shown in the color bar
above the plots.
The polarization data at the \ion{O}{I} and \ion{Ca}{II} lines 
are also shown as a function of Doppler velocity
in Figure \ref{fig:QU_vel}.
\label{fig:QU}}
\end{center}
\end{figure*}

\section{Results}
\label{sec:results}

Figure \ref{fig:pol} shows the observed total flux and 
polarization spectra of SNe 2009jf and 2009mi.
The observed polarization is the sum of the intrinsic 
and interstellar polarization (ISP).
ISP is caused by interstellar dust
both in our Galaxy and in the host galaxy.
However, since the wavelength dependence of the ISP is smooth
\citep{serkowski75},
the changes in the polarization at the strong absorption lines, 
such as \ion{He}{I}, \ion{O}{I}, and \ion{Ca}{II}, 
are certainly intrinsic.
These features clearly indicate a broken symmetry of the SNe.

The same data are shown in the $Q-U$ diagram in Figure \ref{fig:QU}.
Different colors represent different wavelengths, according to
the color bar above the plots.
Most of the data points, except for the strong lines, 
are located around 
($Q_0$, $U_0$) $\simeq$ ($-$0.25 \%, 0.50 \%) and ($-$0.30 \%, 0.60 \%)
for SNe 2009jf and 2009mi, respectively.
Spectropolarimetric type is N1
according to the classification scheme of \citet{wang08}.
Some deviation from the reference point could be due to 
non-zero continuum polarization of SNe.
Hereafter we use these ($Q_0$, $U_0$) as a reference point.
This component is most likely to be the ISP for the following reason.
The ISP can be corrected if we assume a complete depolarization 
at strong emission lines 
\citep[\eg][]{kawabata02,leonard05,wang06,maund0705bf,maund0706aj}.
This method is justified because the emission part of a P Cygni profile 
consists largely of line-scattered, depolarized photons.
In our data, the polarization at the strong emission peak at the 
\ion{Ca}{II} line is consistent with the reference point above.
In the following discussion, however,
we do not explicitly assume this component to be the ISP but
focus only on the line polarization.

The $Q-U$ diagram (Figure \ref{fig:QU}) is a more unambiguous 
way to show the observed polarization data than a plot 
as a function of wavelength (Figure \ref{fig:pol}).
Since the ISP could add an offset to the SN polarization,
the polarization angle $\theta$ in Figure \ref{fig:pol} 
is not directly related to the SN properties.
Instead, one can readily measure the polarization angle 
$\theta'$ from the reference point ($Q_0, U_0$);
\begin{equation}
2\theta' = {\rm atan}[(U-U_0)/(Q-Q_0)].
\end{equation}

\begin{figure*}
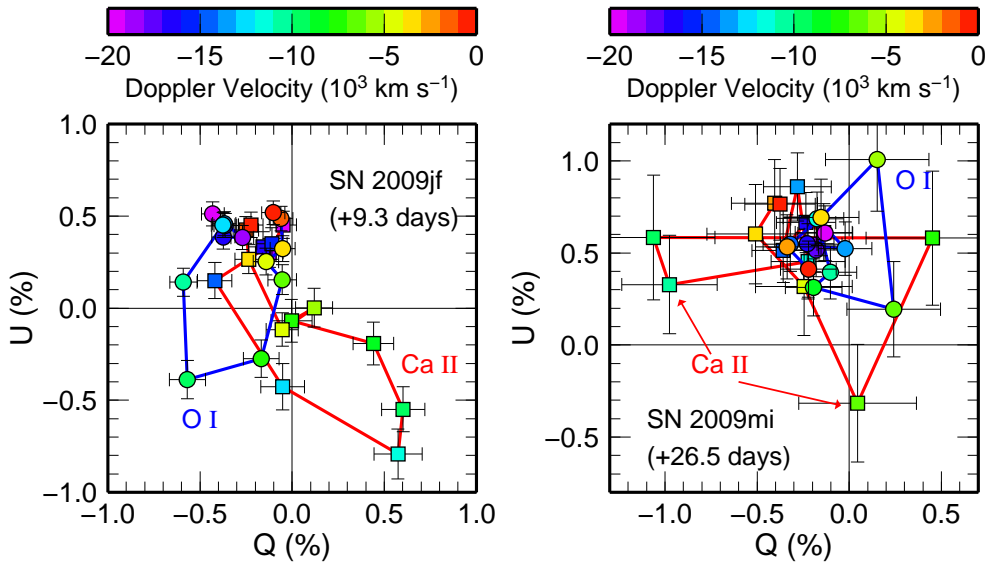

\begin{center}
\begin{tabular}{cc}
\includegraphics[scale=0.70]{f5a.eps} &
\includegraphics[scale=0.70]{f5b.eps}
\end{tabular}
\caption{
Observed polarization data of SNe 2009jf (left) and 2009mi (right)
around the \ion{O}{I} line (circles connected with the blue line) 
and \ion{Ca}{II} line (squares connected with the red line).
The polarization data are binned into 50 \AA, giving 
a velocity resolution of about 1900 and 1700 \kms\ for the \ion{O}{I} and
\ion{Ca}{II} lines, respectively.
The interstellar polarization is {\it not} corrected for.
Different colors show different Doppler velocities
as shown in the color bar above the plots.
Although the \ion{Ca}{ii} triplet seems to be resolved
in the total flux of SNe 2009jf and 2009mi (Figure \ref{fig:pol}),
the velocity in this Figure is simply measured from the mean 
wavelength (8567 \AA).
In our data, there is no evidence for an additional high-velocity 
component of the \ion{Ca}{ii} line as seen in some
other SNe (see \eg \citealt{maund09}).
The \ion{O}{i} and \ion{Ca}{ii} features show 
a loop in the $Q-U$ plane, indicating 
non-axisymmetric distribution.
In addition, the \ion{Ca}{II} line in SN 2009mi shows a large change of the 
angle measured from the reference point.
\label{fig:QU_vel}}
\end{center}
\end{figure*}

In Figure \ref{fig:QU_vel}, we show the data
only around the \ion{O}{I} and \ion{Ca}{II} lines
as a function of Doppler velocity.
It is clear that
the data at the \ion{Ca}{II} and \ion{O}{I} lines occupy different regions 
in the $Q-U$ diagram, indicating different spatial distributions between
\ion{Ca}{II} and \ion{O}{I}.
Such a difference is also clearly seen in 
other SNe \eg 
Type Ib SN 2008D \citep{maund09} and Type IIb SN 2008ax \citep{chornock11}.

A more interesting feature is the shape of the polarization 
data in the $Q-U$ diagram.
Starting from the reference point, the \ion{Ca}{II} and \ion{O}{I} lines
in SN 2009jf show a loop at these lines
(spectropolarimetric type L).
This means that the angle $\theta'$ measured from the reference 
varies with Doppler velocity, and the depth in the ejecta 
(homologous expansion, $r = vt$).
As shown in Figure \ref{fig:intro_sim}, and 
as suggested by \eg \citet{kasen03,maund0705bf,maund0701ig},
this loop clearly indicates that even axisymmetry is broken
in the SN ejecta.

The \ion{Ca}{II} feature of SN 2009mi is even more intriguing.
Measuring from the reference point, the variation in
the angle $\theta'$ is as large as $\sim 90^{\circ}$
(the difference of $180^{\circ}$ in the $Q-U$ diagram corresponds
to the difference of $90^{\circ}$ in the polarization angle on the sky).
This indicates a large change in the distribution of the \ion{Ca}{II} ion
depending on the depth of the ejecta.

\section{Properties of Supernova Line Polarization}
\label{sec:properties}

\subsection{Ubiquity of Non-axisymmetry}

In Table \ref{tab:pol}, 
we summarize line polarization measurements obtained so far 
for stripped-envelope SNe.
Here the line polarization is defined as a relative change 
from the continuum level, and thus, not affected by the ISP.
We show 6 stripped-envelope SNe with high-quality data,
namely, Type Ib SNe 2005bf \citep{maund0705bf,tanaka0905bf}, 
2008D \citep{maund09}, 2009jf (this paper), Type Ic SNe 2002ap 
\citep{kawabata02,leonard0202ap, wang0302ap}, 2007gr \citep{tanaka0807gr},
and 2009mi (this paper).
As also noted by \citet{wang08}, it is clear that all SNe show non-zero polarization, 
which means that stripped-envelope SNe generally have asymmetric explosion geometry.

For SNe 2005bf and 2008D, \citet{maund0705bf,maund09} found
a loop in the $Q-U$ diagram at strong lines,
which is indicative of a 3D geometry
\citep[see also Figure \ref{fig:intro_sim}]{kasen03,maund0705bf,maund0701ig}.
Our new data for SNe 2009jf and 2009mi 
also show a loop in the $Q-U$ diagram.
We have checked the literature about SN 2002ap 
\citep{kawabata02,leonard0202ap, wang0302ap}.
Although they do not explicitly mention the loop,
the data in \citet{kawabata02} show a loop in the \ion{Ca}{ii} line.
Thus, loops are quite common in stripped-envelope SNe;
five of six stripped-envelope SNe show the loop.
This implies that a non-axisymmetric, 
3D geometry is common in stripped-envelope SNe.

\subsection{Relation Between Line Polarization and 
Absorption Depth}

\begin{figure}
\begin{center}
\includegraphics[scale=0.70]{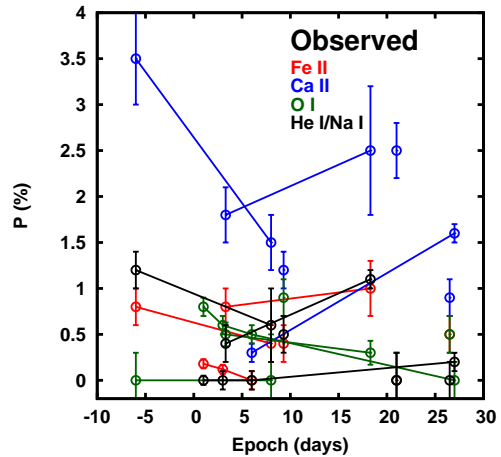}
\caption{
Observed line polarization as a function of SN epoch.
Data for the same SN are connected with lines.
There is no clear trend in the time evolution of 
the line polarization.
The \ion{Ca}{ii} line tends to be more polarized than other lines.
\label{fig:Pobs_epoch}}
\end{center}
\end{figure}

Figure \ref{fig:Pobs_epoch} shows the observed line polarization
as a function of epoch from maximum brightness.
There is a large variety in the polarization degree.
It is clear that the \ion{Ca}{ii} line (blue) 
tends to be more polarized than the other lines.
There is no clear trend in the time evolution;
the line polarization could increase or decrease with time.

\begin{figure}
\begin{center}
\includegraphics[scale=0.70]{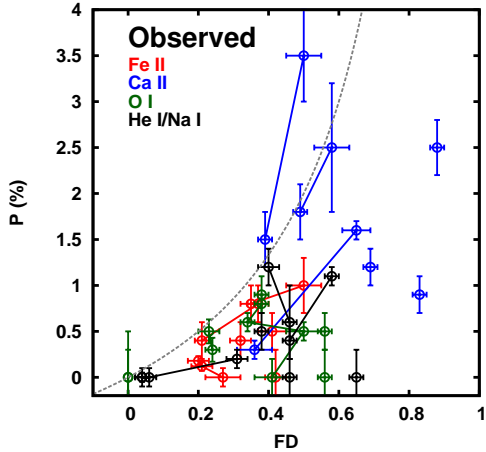}
\caption{
Observed line polarization as a function of fractional depth
of the absorption.
Data for the same SN are connected with lines.
There is a general trend that stronger lines show higher polarization,
as expected from Equation \ref{eq:Pabs}.
The gray dashed line shows the line polarization as a function of 
FD which is expected by Equation \ref{eq:Pabs} 
for the case of $P_{\rm cor} = 2.0 \%$.
\label{fig:Pobs_FD}}
\end{center}
\end{figure}

Figure \ref{fig:Pobs_FD} shows the same polarization
data as a function of the fractional depth of the absorption. 
The fractional depth (FD) is defined as
\begin{equation}
{\rm FD} = \frac{F_{\rm cont} - F_{\rm abs}}{F_{\rm cont}},
\label{eq:FD}
\end{equation}
where $F_{\rm cont}$ and $F_{\rm abs}$ are the flux at the 
continuum and absorption minimum, respectively.

Figure \ref{fig:Pobs_FD} shows that 
stronger lines tend to show higher polarization degree.
In fact, there are no observational points at 
the top left of the plot, \ie FD $< 0.3$ and $P_{\rm obs} > 1 \%$.
This seems natural because the absorption, unless it is quite strong, 
cannot cause a significant incomplete cancellation of the polarization
(Figure \ref{fig:ejecta}).

\begin{figure*}
\begin{center}
\begin{tabular}{cc}
\includegraphics[scale=0.70]{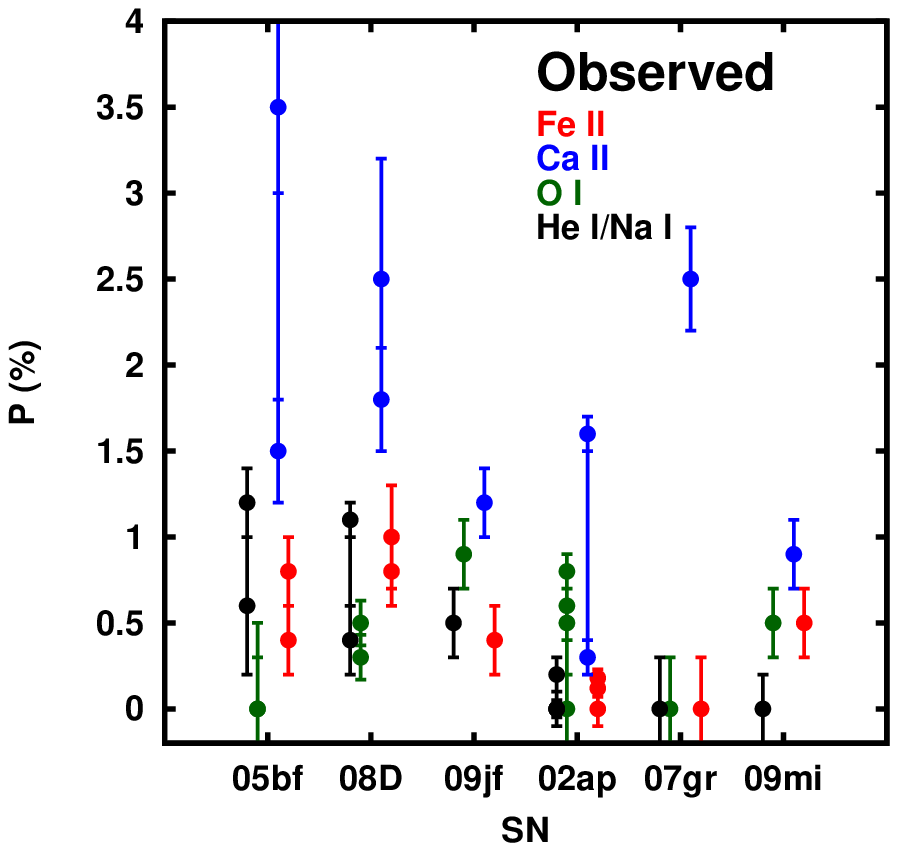} &
\includegraphics[scale=0.70]{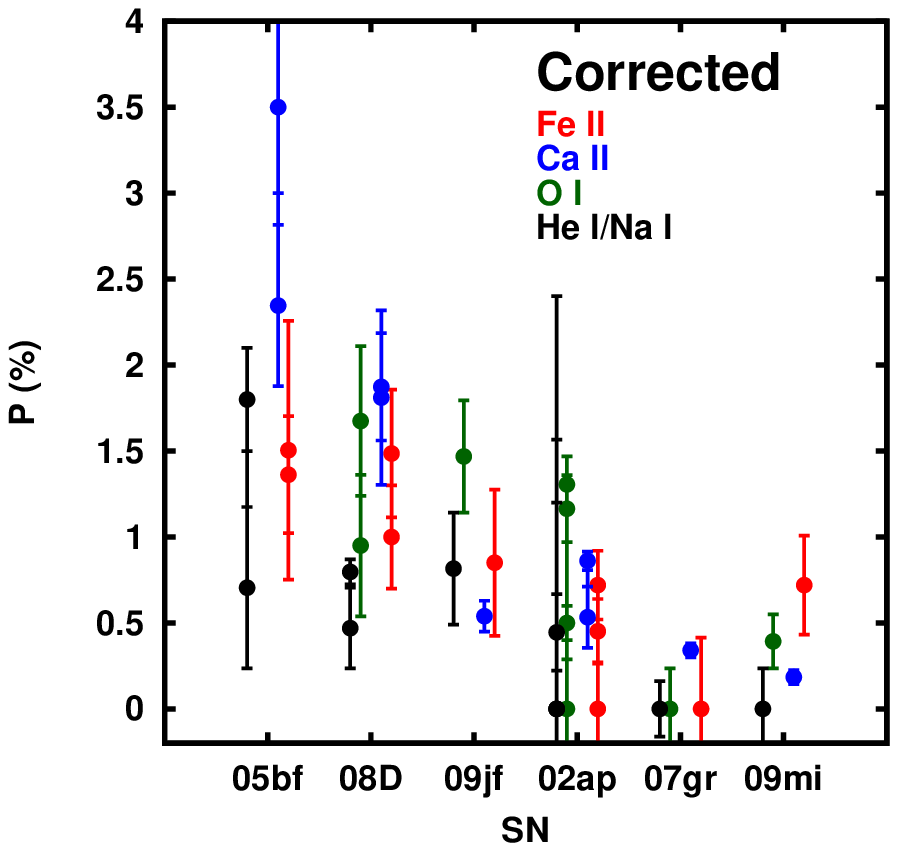}
\end{tabular}
\caption{
({\it Left}) Observed line polarization for stripped-envelope SNe.
The \ion{Ca}{ii} line is usually more polarized than the other lines.
({\it Right}) Line polarization corrected for the absorption depth
(Equation \ref{eq:Pabs}).
After correcting the effect of the absorption depth, 
the \ion{Ca}{ii} lines is not always more polarized than the other lines.
\label{fig:Pobs_SN}}
\end{center}
\end{figure*}

We can analyze this relation in a simple, analytic way.
We assume a spherical photosphere projecting an area $S$ on the sky.
Within this photospheric disk, the flux per unit area $I$
and the polarization degree are assumed to be uniform.
The total flux from the photospheric disk is $F_{\rm cont} = IS$.
When light is emitted from a point on the photospheric disk
on the sky, it can undergo line absorption with an
absorption fraction $x_{\rm abs} \equiv 1 - \exp(-\tau)$.
At the wavelength of this line, the flux is 
$F_{\rm abs} = (1-x_{\rm abs})IS$.

If the distribution of the line opacity is uniform,
no polarization is expected at the line, \ie $P_{\rm abs} = 0$
(Section \ref{sec:intro_specpol}).
To introduce asymmetry in the line opacity, 
we assume that the absorption fraction is enhanced 
by a factor of $f$ in a region $\Delta S$
(\eg the shaded region in Figure 1(c)).
Then, the total flux at the wavelength of the line is 
\begin{eqnarray}
F_{\rm abs} &=& (1-x_{\rm abs}) I (S-\Delta S) + (1-fx_{\rm abs}) I \Delta S \nonumber \\
&=& [(1-x_{\rm abs}) - (f-1)x_{\rm abs} \Delta S/S] IS,
\label{eq:Fabs}
\end{eqnarray}
and the fractional depth (Equation \ref{eq:FD}) 
can be written by using Equation \ref{eq:Fabs}
\begin{equation}
{\rm FD} = x_{\rm abs} + (f-1)x_{\rm abs} \Delta S/S.
\label{eq:FD2}
\end{equation}

The asymmetry introduced above results in an 
incomplete cancellation of the polarization.
If this region has a constant polarization direction, 
the amount of non-cancelled flux is equivalent to 
the excess absorption in the region, \ie 
\begin{equation}
P_{\rm abs}F_{\rm abs} = (f-1)x_{\rm abs} I \Delta S.
\label{eq:PFabs}
\end{equation}
Then, the polarization degree can be expressed as a function of 
the fractional depth by Equations \ref{eq:FD2} and \ref{eq:PFabs}:
\begin{equation}
P_{\rm abs} = \frac{(f-1)x_{\rm abs} \Delta S/S}{1 - {\rm FD}}
\end{equation}
Unless the enhanced opacity dominates the absorption, 
${\rm FD} \simeq x_{\rm abs}$ (Equation \ref{eq:FD2}).
Thus, 
\begin{equation}
P_{\rm abs} \simeq (f-1)\Delta S/S \frac{{\rm FD}}{1 - {\rm FD}} 
= P_{\rm cor} \frac{{\rm FD}}{1 - {\rm FD}},
\label{eq:Pabs}
\end{equation}
where we define a corrected polarization degree 
$P_{\rm cor} \equiv (f-1)\Delta S/S$.
This corrected polarization degree is equivalent 
to the polarization with ${\rm FD}=0.5$.
It can be interpreted as a rough indicator of the product of 
the amount of the enhancement $(f-1)$ and the 
fractional area of the enhanced region $\Delta S/S$ under the assumption that 
the enhanced region only has one polarization direction.
The gray dashed line in Figure \ref{fig:Pobs_FD} shows 
the line polarization expected by Equation \ref{eq:Pabs} 
with $P_{\rm cor} = 2.0 \%$.

Note that the analytic equations here involve
significant simplifications, \ie a constant intensity 
in the photospheric disk, and a constant direction of 
polarization in the enhanced region.
But if we use $\Delta S$ as an intensity-weighted area, 
in which polarization is not cancelled, 
the equation holds even for general cases with
a non-constant intensity and many opacity-enhanced regions.

From Equation \ref{eq:Pabs}, it is expected that 
stronger lines with a larger fractional depth tend to have higher polarization, 
which is actually found in Figure \ref{fig:Pobs_FD}.
It should be emphasized that the variety 
of the observed line polarization is largely generated 
by a variety of the absorption depth.
The absorption depth is primarily determined by the density, 
chemical abundance, or ionization in the ejecta,
which can be varied even in the 1D explosion
(reflecting the mass and kinetic energy of the ejecta, 
and the temperature in the ejecta).

Figure \ref{fig:Pobs_SN} shows
the observed polarization (left) and polarization corrected for 
the absorption depth (right) for each stripped-envelope SN.
It is obvious that after correction for the absorption depth,
the \ion{Ca}{ii} line is not always more polarized than the other lines.
This demonstrates the importance of the correction. 
To compare the polarization of different lines or different objects correctly, 
one must correct for this effect, which is not directly related to the geometry.
This fact has been overlooked in previous analyses.
It would be interesting to study this effect also on the
line polarization of Type Ia SNe,
which has been suggested to be correlated with 
the declining rate of the light curve \citep{wang07}.

\begin{figure*}
\begin{center}
\begin{tabular}{cc}
\includegraphics[scale=0.70]{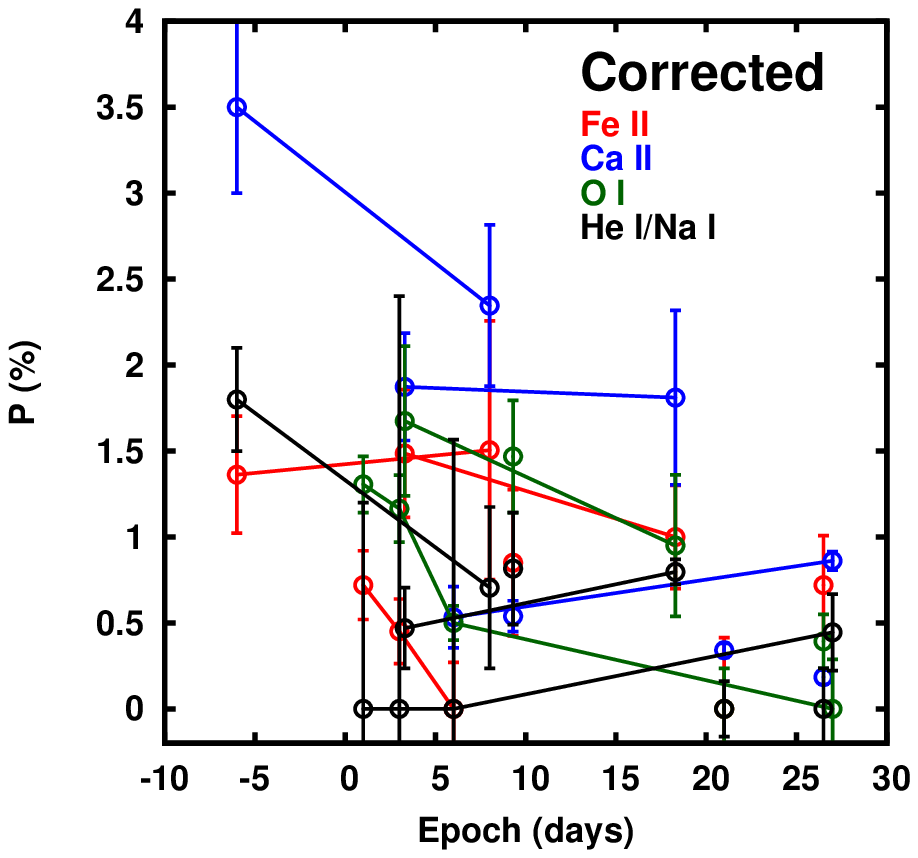} &
\includegraphics[scale=0.70]{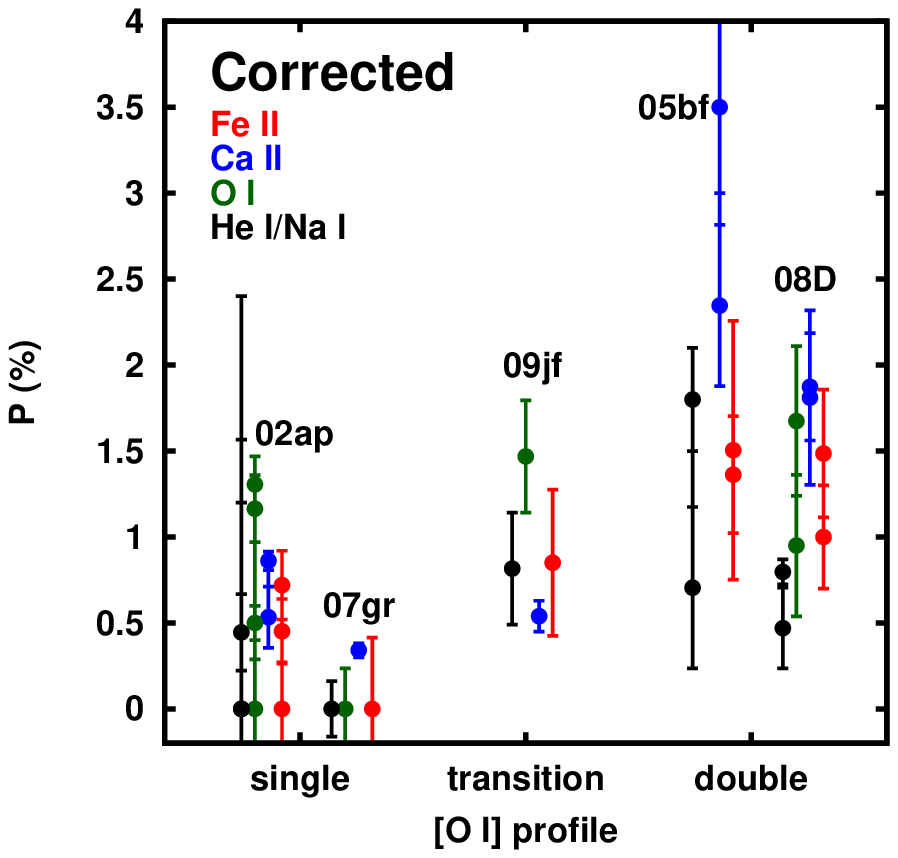}
\end{tabular}
\caption{
({\it Left}) Corrected line polarization as a function of SN epoch.
Data for the same SN are connected with lines.
({\it Right}) Relation between corrected line polarization 
and nebular [\ion{O}{i}] line profile.
\label{fig:Pcor}}
\end{center}
\end{figure*}

\subsection{Properties of Corrected Polarization}

It is clear that there is a remaining dispersion
in the line polarization among different objects 
even after the correction for the absorption depth 
(right panel in Figure \ref{fig:Pobs_SN}).
We first examine if the corrected polarization is related with the epoch.
The left panel of Figure \ref{fig:Pcor} shows 
the corrected line polarization as a function of epoch.
Although there is a weak trend that the polarization decreases with time,
it is not clear if all the variety in the corrected polarization 
can be explained by the effect of the epoch.

The remaining dispersion could also be understood 
by the multi-dimensional properties of each SN.
It is thus worth studying a relation between 
the line polarization and 
the [\ion{O}{i}] line profile in the nebular spectra,
which is another geometric probe of the SN ejecta
\citep{mazzali01,maeda02,mazzali05,maeda08,modjaz08,
tanaka0908Dneb,taubenberger09,maurer10}.

The right panel of Figure \ref{fig:Pcor} shows 
the relation between corrected line polarization and 
the [\ion{O}{i}] line profile.
SNe 2002ap and 2007gr show a single-peak profile 
\citep{leonard0202ap,foley03,mazzali07,hunter09,mazzali10}
while SNe 2005bf and 2008D show a double-peak profile 
\citep{maeda07,modjaz09,tanaka0908Dneb}.
The [\ion{O}{i}] profile of SN 2009jf is complex; 
it shows both single- and double-peak components \citep{sahu11,valenti11}.
Following the classification by \citet{maeda08}, 
we classify it as ``transition''.
Only with five objects, 
it is difficult to draw a firm conclusion.
It is therefore important to increase spectropolarimetric samples
to further study this relation.

The dispersion may also reflect a difference between Type Ib and Ic.
In fact, 3 Type Ic SNe in our sample tend to have 
lower polarization degree than 3 Type Ib SNe
(left panel of Figure \ref{fig:Pobs_SN}).
By increasing the number of samples, 
we will be able to study the effect of the existence of the He envelope 
on the explosion geometry.

One natural scenario is that 
the dispersion is caused by random line-of-sight effects.
As discussed in Section \ref{sec:results},
stripped-envelope SNe generally have 3D geometry.
If we assume a clumpy structure as in the left bottom
panel in Figure \ref{fig:intro_sim},
a variety of polarization is naturally expected
for different line-of-sight.
Since the effect of the line-of-sight is very sensitive
to the size and the number of clumps, 
it can be used as an observational probe
to study a typical 3D geometry of SNe.
This will be further studied in a forthcoming paper.

\section{Conclusions}
\label{sec:conclusions}

We have performed spectropolarimetric observations of
2 stripped-envelope SNe, Type Ib SN 2009jf and Type Ic SN 2009mi.
Both objects show non-zero line polarization at 
the wavelength of strong absorption lines.
The polarization data show a "loop"
in the $Q-U$ diagram, which indicates 
a non-axisymmetric, 3D distribution of ions in the SN ejecta.
After adding our new data to the sample of stripped-envelope SNe with
high-quality spectropolarimetric observations,
five SNe out of six show a loop in the $Q-U$ diagram.
We conclude that a non-axisymmetric, 3D geometry 
is common in stripped-envelope SNe.

We have studied the properties of line polarization 
in stripped-envelope SNe.
We have found that a stronger line tends to show 
a higher polarization.
We give an analytic equation that naturally 
explains this relation.
Although the \ion{Ca}{ii} line usually shows a higher 
line polarization than the other lines, 
it is not always true after the correction of the absorption depth.
This effect must be corrected in order to compare the polarization of 
different lines or different objects.

Even after the correction for the absorption depth, 
there is a dispersion in the polarization degree among different objects.
This dispersion could simply reflect the epoch of the observations.
It might also be related to 
the [\ion{O}{i}] line profile in the nebular spectra
or SN types (Type Ib or Ic), 
although it is not possible to draw any firm conclusion 
with the current sample.
Also, if we assume a 3D clumpy structure, 
the variety is naturally expected by the line-of-sight effect.
Modeling of spectropolarimetric data will be performed 
in a forthcoming paper.

\acknowledgments
We are grateful to the staff and observers of the Subaru Telescope,
especially Masahiko Hayashi and Hiroshi Terada
for their effort on the time allocation of our ToO observations in 2009.
MT thanks Yudai Suwa and Takami Kuroda for valuable comments.
This research has been supported 
by the Grant-in-Aid for Scientific Research of the 
Japan Society for the Promotion of Science (22840009, 24740117),
and by World Premier International Research Center Initiative, MEXT, Japan, 
This research has made use of the SUSPECT
\footnote{http://suspect.nhn.ou.edu/\~{}suspect/}, 
the Online Supernova Spectrum Archive
at the Department of Physics and Astronomy, University of Oklahoma.


\clearpage

\begin{landscape}

\begin{deluxetable*}{lcccccccccccc} 
\tabletypesize{\scriptsize}
\tablewidth{0pt}
\tablecaption{Summary of Line Polarization}
\tablehead{
Object & Type & 3D?   &  Epoch & $P_{\rm FeII}$ & $P_{\rm CaII}$ & $P_{\rm OI}$ & $P_{\rm NaI/HeI}$ & FD$_{\rm FeII}$ & FD$_{\rm CaII}$ & FD$_{\rm OI}$ & FD$_{\rm NaI/HeI}$ & Ref. \\
       &      &       &  (day) &  (\%)        & (\%)         &  (\%)       &  (\%)          &               &               &             &                  &
}
\startdata
SN 2005bf   &  Ib   & yes &    -6    &  0.8 (0.2)  &  3.5 (0.5)  & 0.0 (0.3) &  1.2 (0.2) & 0.37 (0.03) & 0.50 (0.05) & 0.0 (0.0)   & 0.40 (0.03) & 1 \\
            &       &     &    8     &  0.4 (0.2)  &  1.5 (0.3)  & 0.0 (0.5) &  0.6 (0.4) & 0.21 (0.02) & 0.39 (0.02) & 0.0 (0.0)   & 0.46 (0.02) & 2 \\
SN 2008D    &  Ib   & yes &    3.3   &  0.8 (0.2)  &  1.8 (0.3)  & 0.5 (0.13)&  0.4 (0.2) & 0.35 (0.03) & 0.49 (0.02) & 0.23(0.03)  & 0.46 (0.02) & 3 \\
            &       &     &   18.3   &  1.0 (0.3)  &  2.5 (0.7)  & 0.3 (0.13)&  1.1 (0.1) & 0.50 (0.05) & 0.58 (0.05) & 0.24 (0.02) & 0.58 (0.02) & 3 \\
SN 2009jf   &  Ib   & yes &    9.3   &  0.4 (0.2)  &  1.2 (0.2)  & 0.9 (0.2) &  0.5 (0.2) & 0.32 (0.03) & 0.69 (0.02) & 0.38 (0.02) & 0.38 (0.02) & this paper \\ \hline
SN 2002ap   &  Ic   & yes &    1     &  0.18 (0.05)&   --        & 0.8 (0.1) &  0.0 (0.05)& 0.20 (0.03) &    --       & 0.38 (0.02) & 0.04 (0.02) & 4,5 \\
            &       &     &    3     &  0.12 (0.05)&    --       & 0.6 (0.1) &  0.0 (0.1) & 0.21 (0.03) &    --       & 0.34 (0.02) & 0.04 (0.02) & 4,5 \\
            &       &     &    6     &  0.0  (0.1) & 0.3 (0.1)   & 0.5 (0.1) &  0.0 (0.1) & 0.27 (0.05) & 0.36 (0.05) & 0.50 (0.05) & 0.06 (0.02) & 6 \\
            &       &     &    27    &    --       & 1.6 (0.1)   & 0.0 (0.2) &  0.2 (0.1) &  --         & 0.65 (0.04) & 0.41 (0.05) & 0.31 (0.03) & 4 \\
SN 2007gr   &  Ic   & no  &    21    &  0.0 (0.3)  & 2.5  (0.3)  & 0.0 (0.3) &  0.0 (0.3) & 0.42 (0.03) & 0.88 (0.02) & 0.56 (0.02) & 0.65 (0.02) & 7 \\
SN 2009mi   &  Ic   & yes &    26.5  &  0.5 (0.2)  & 0.9  (0.2)  & 0.5 (0.2) &  0.0  (0.2 & 0.41 (0.04) & 0.83 (0.02) & 0.56 (0.02) & 0.46 (0.02) & this paper \\ 
\enddata
\tablerefs{(1) \citealt{maund0705bf}, (2) \citealt{tanaka0905bf},
(3) \citealt{maund09}, (4) \citealt{kawabata02}, (5) \citealt{wang0302ap}, (6) \citealt{leonard0202ap},
(7) \citealt{tanaka0807gr}
}
\label{tab:pol}
\end{deluxetable*}
\clearpage

\end{landscape}

\end{document}